# Polymer amide as a source of the cosmic 6.2 micron emission and absorption

Julie E. M. McGeoch[1] and [2]Malcolm W. McGeoch

1. High Energy Physics DIV, Smithsonian Astrophysical Observatory Center for Astrophysics | Harvard & Smithsonian, 60 Garden St, MS 70, Cambridge MA 02138, USA.

2. PLEX Corporation, 275 Martine St., Suite 100, Fall River, MA 02723, USA.

**Abstract**
Cosmic infrared emission and absorption spectra often carry a well-defined and invariant 6.2 micron band that has been proposed to emanate from very small dust grains that may carry polyaromatic hydrocarbons. Hemoglycin, a well-defined polymer of glycine that also contains iron, has been found in meteorites of the primordial CV3 class and therefore originated in the solar protoplanetary disc. In approximate calculations, the principal amide I infrared absorption band of hemoglycin is at 6.04 microns. Hemoglycin, an antiparallel beta sheet structure with two 11-mer glycine chains, has an exact structural analog in antiparallel poly-L-lysine beta sheets which in the laboratory have an absorption peak at 6.21 microns. This wavelength coincidence, the demonstrated propensity of hemoglycin 4.9nm rods to form accreting lattice structures, and its proven existence in the solar protoplanetary disc strongly suggest that the cosmic 6.2 micron emission and absorption could be from small grains that are hemoglycin lattices or shell-like vesicles carrying internal organic molecules of various types. Calculated hemoglycin ultraviolet absorptions associated with iron in the molecule match the observed ultraviolet extinction feature at nominal 2175 Angstroms.

**Main text**
Infrared emission spectra from celestial objects often carry a relatively sharp 6.2 μm band that at moderate resolution can be the most intense molecular feature in the 3 – 12 micron range [1]. Recent JWST data (Berne et al 2023 [2], and private communication [3]) show this band at a very high resolution, instanced by the atomic and ionic lines included (Figure 1). Infrared molecular emissions in general have been widely attributed to poly aromatic hydrocarbons (PAHs) although questions persist about the identity of the prominent 6.2 μm emission [4,5]. The 6.2 μm band and other emissions between 6.2μm and 9.0 μm have been grouped as coming from carbonaceous very small grains (VSGs) in an analysis by Foschino et al. [1] who detected a commonality of behavior at 6.2, 7.6 and 7.8 μm, displaying this "eVSG" component separately in a detailed breakdown of 31 high resolution IR spectra covering 2.5 – 15 μm. Variations within the group were attributed to varying degrees of PAH ionization, or evaporation from these grains, that had an estimated temperature of less than 270K [5,6].

Hemoglycin [7] was first identified [8,9] in extracts of class CV3 carbonaceous meteorites and consists of a pair of antiparallel chains each of 11 glycine residues closed out at the ends by iron atoms (Figure 2, reproduced from [7]). Its calculated infrared absorption spectrum in vacuo [7], within an accuracy of about 5%, has a very prominent amide I band at approximately 1656 cm$^{-1}$ (6.04 μm). More emphasis should be laid on the laboratory measurement of a similar anti-parallel beta sheet structure, poly-L-lysine, via Fourier Transform Infrared (FTIR) spectroscopy [10] which reveals a very strong absorption peak at 1611 cm$^{-1}$ (6.21 μm) and a weaker peak at 1680cm$^{-1}$ (5.95 μm). These relate to the "α-" and "α+" transitions [10], respectively, formed by a splitting of the amide I band in this protein configuration. Antiparallel $\beta$-sheet amide I absorption and emission strength is exceptional in comparison to its molecular weight because a linear array of N aligned and identical radiating dipoles, one per residue, radiates power proportional to N$^2$. The dipoles of PAH molecules, of lesser molecular weight, are not linearly aligned and should proportionately radiate much less strongly.

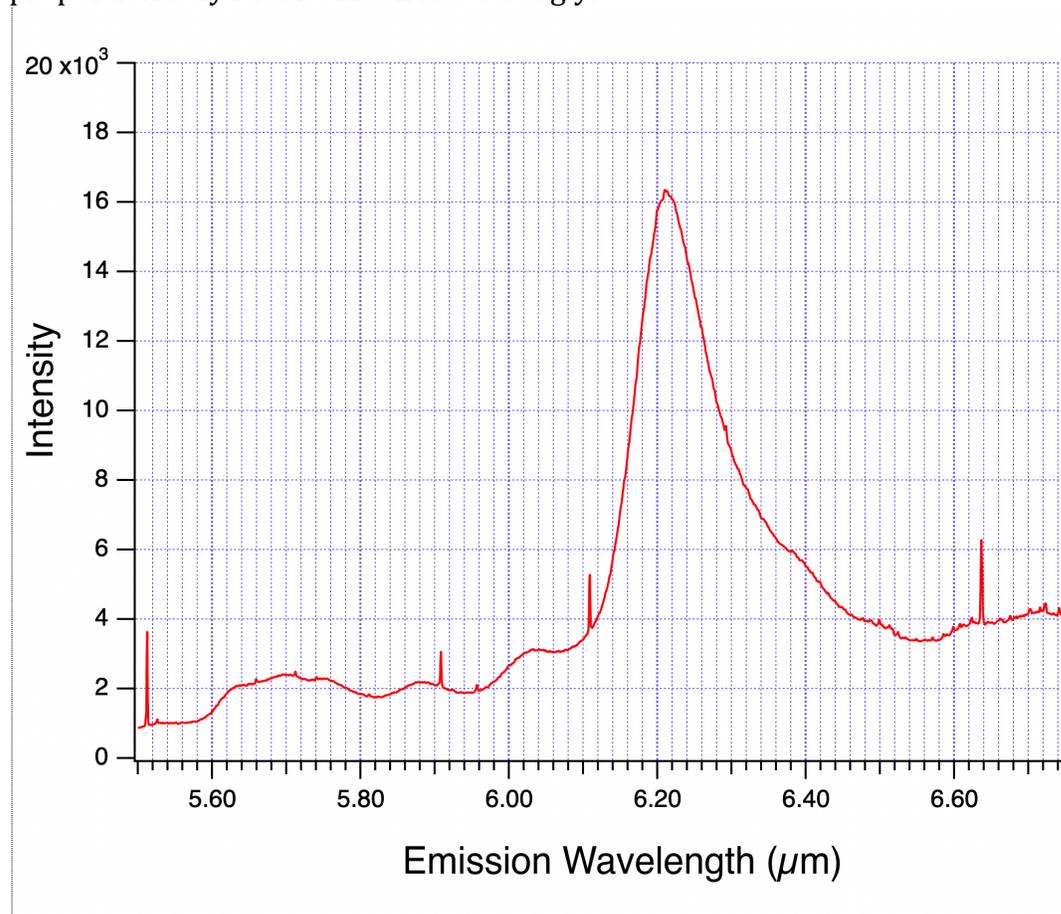

**Figure 1 Emission from the JWST MIRI channel 1 spectrum of the d203-506 protoplanetary disk showing the prominent 6.21μm peak.** The data was provided by Olivier Berné et al to J.E.M.Mc [3] and is available via O Berné et al. Formation of the Methyl Cation by Photochemistry in a Protoplanetary Disk. Nature https://doi.org/10.1038/s41586-023-06307-x (2023).

Possibly because individual amino acid residues have not been observed to date in molecular clouds or even in protoplanetary discs, the existence in space of polymers of amino acids has not been considered seriously. In any case the imagined complexity of polymer amide would seemingly present an insurmountable barrier to molecular identification. However, it was shown calculationally [11] that exothermic polymerization of amino acids could occur slowly in the conditions of warm, dense molecular clouds provided the water of condensation remained attached by hydrogen bonds to the product polymer. Additionally, hemoglycin has been found in a range of mainly class CV3 meteorites that are relatively unaltered and can be assumed to carry vestiges of the earliest accreted material in our own protoplanetary disc [12]. Hemoglycin carries high $^{2}$H and $^{15}$N content [9,13] that could be explained by a molecular cloud origin, making it plausible to now think in terms of a polymer amide content in interstellar space.

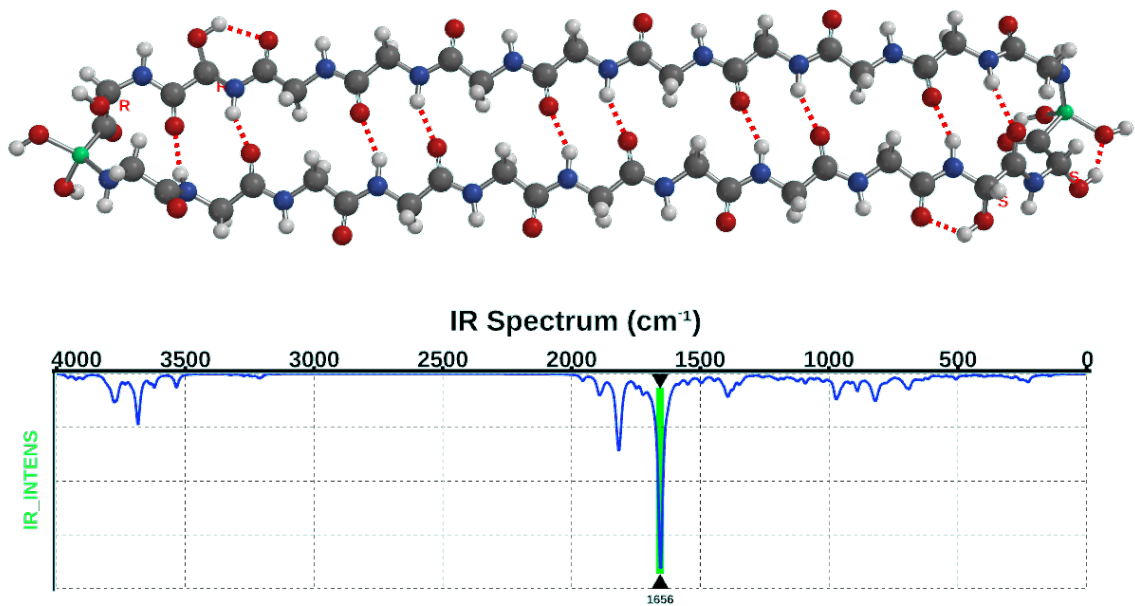

**Figure 2. (reproduced from [7]). Calculated IR absorption from the core of Glycine$_{18}$ Hydroxy-glycine$_4$ Fe$_2$O$_4$ shown above. An absorption peak near 6 μm shown here at 1656cm$^{-1}$, is the "amide I" absorption for the backbone of the anti-parallel beta sheet. Molecule format is ball and spoke. Atom labels: hydrogen white, carbon black, nitrogen blue, oxygen red, iron green.**

In our calculations [7], which are only accurate in an absolute sense to 5% in wavelength, there are several different extended vibrational modes within the hemoglycin amide I absorption, ranging from 1586cm$^{-1}$ (6.30 μm) to 1724cm$^{-1}$ (5.80 μm), the calculated full base width of the band being about 0.5 μm. The corresponding infrared <u>emission</u> could be slightly red-shifted owing to rapid (sub-picosecond) intra-molecular migration of absorbed energy through modes of nearly the same energy [14]. A cascade of internal transitions through these closely spaced vibrational modes could contribute to longer wavelength emissions that extend by up to 0.2 μm from the main emission peak.

In the IR emission spectra between 2.5 and 15 μm listed in Appendix D of [1] there is associated with the dominant 6.2 μm emission an uneven continuum of emissions growing toward 7.6/7.8 μm. At the same time, hemoglycin should only have an isolated 6.2-6.3μm emission in that region, without significant additional features between 6.3μm and 8μm (Figure 2). It is therefore suggested that accreted contents within small fragments of three-dimensional hemoglycin lattice (Figure 3) could be responsible for the characteristic spectral content between 6.3 and 9 μm included in [1] as emission related to "very small grains" (VSGs). An emission event will most likely begin with absorption of a short-wavelength photon followed by visible radiation that populates vibrational levels in either the polymer lattice or its accreted contents. Before the complex loses this energy it can be passed internally in either direction between hemoglycin lattice polymer rods and the accreted contents. An exchange of energy can explain the documented tracking [1] of 7.6/7.8 μm emissions with the 6.2μm emission.

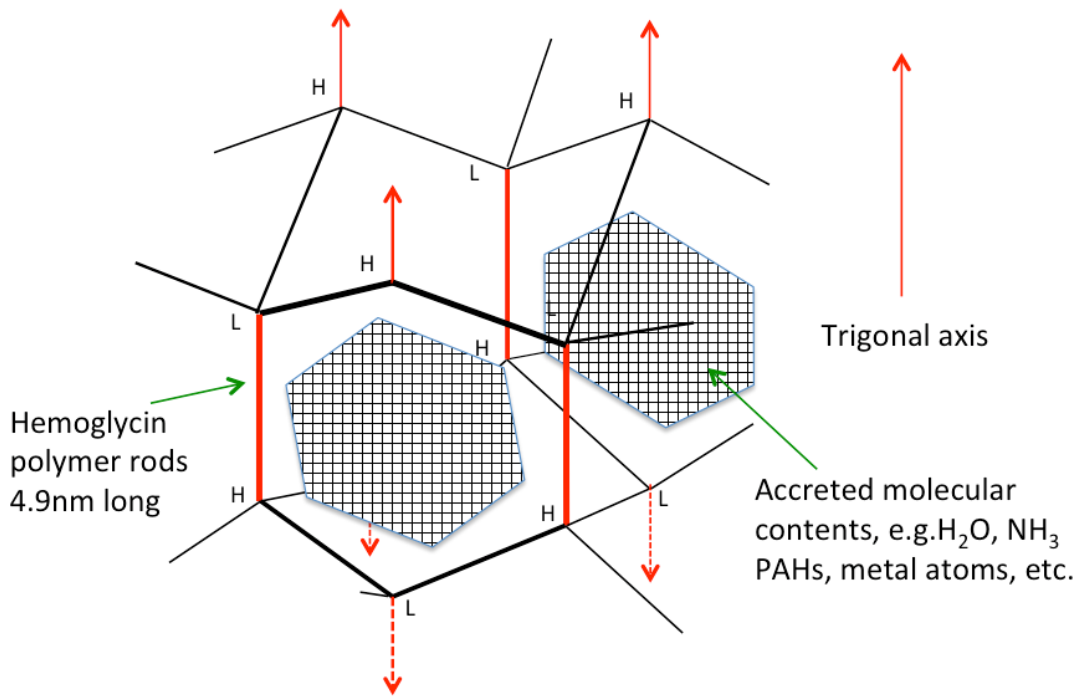

**Figure 3. Three-dimensional hemoglycin lattice with diamond 2H structure, showing accreted contents. Lattice plus contents are a candidate for the "very small grain" 6 – 8 micron emissions. High and low lattice vertices are labeled in the lattice description of [15].**

In relation to accretion we have shown that hemoglycin can form three-dimensional lattices [15,16]. X-ray studies of one crystalline meteoritic extract containing a lattice showed nickel nano-crystals within the lattice that had a dimension of 8±1 nm via

Scherrer analysis, which fits within the lattice cell volume. Within another example of the lattice we have observed calcium carbonate crystals [16]. It therefore is reasonable to generalize this behavior to the collection by hemoglycin lattice fragments in molecular clouds (or protoplanetary discs) of a wide range of substances. The resulting entities would appear observationally to be dust grains that had various IR emissions related to the accreted contents together with a dominant ever-present emission at 6.2µm.

Interstellar absorption at 6.2microns has been presented by Schutte et al [17] via the background lighting of galactic center sources. In [17, Fig.5] two spectra, from GCS3 and GCS4, have very sharp absorption features at 6.22 and 6.18 microns, respectively. These spectra may also show evidence of the 5.95 micron α+ minor peak of antiparallel beta sheet absorption [10] (Figure 4). In future comparisons with laboratory data astronomical absorption measurements will be much more valuable than emission as they are not affected by the intramolecular transitions that can precede emission.

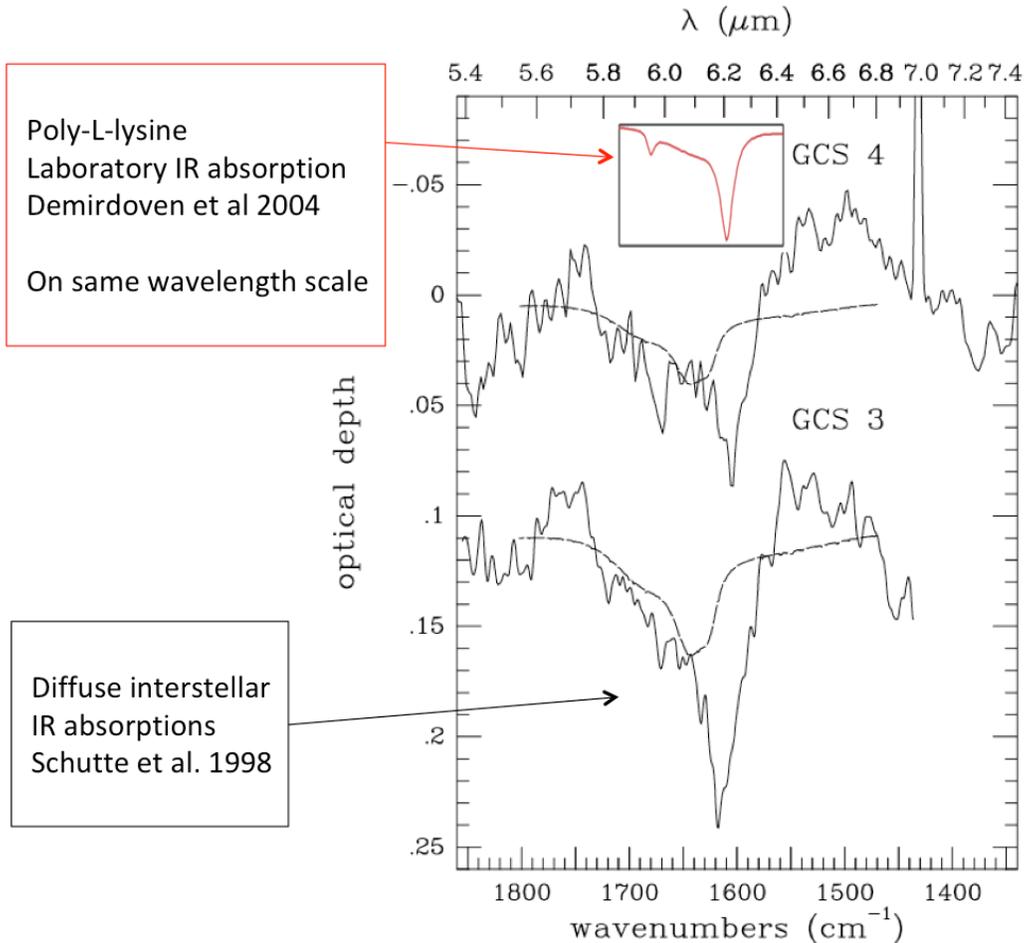

**Figure 4. Main image: Figure 5 from Schutte et al. [17]. Superimposed with the same wavelength scale is the absorption of poly-L-lysine measured in the laboratory by Demirdoven et al. [10]. The smooth curve is an estimated ice mixture absorption calculated by Schutte et al [17] that may also be present.**

A further aspect of hemoglycin spectroscopy, its ultraviolet absorption, may possibly relate to the 2175 Angstrom extinction "bump" associated with interstellar dust [ref. 18, Appendix A1, footnote 11]. This feature has been attributed to PAH absorption, but it is noted in [18] that the interstellar feature is narrower than experimental absorption spectra from mixtures of PAHs. Observational evidence now couples the mid infrared emissions from 6 – 8 microns more definitely with the 2175 Angstrom bump [19], at the same time raising further questions about the real identity of the common molecule responsible for both phenomena. Wavelength variation of the 2175 Angstrom is described [19] of the 2175 Angstrom bump out to a central value of 2350 Angstrom. In the 2000 to 2400 Angstrom region the strongest calculated hemoglycin absorptions are as follows [ref. 7, Table 2]: 2250A, strength 0.041; 2280A, strength 0.062; 2400A, strength 0.141.

In a new observation [20] at high redshift (z=6.71, corresponding to about 800My into cosmic time) the bump is centered at 2263 Angstroms (+20A, -24A), which centrally matches the pair of hemoglycin absorptions at 2250 and 2280 Angstroms, and includes within its overall width the 2400Angstrom band. These are calculated absorption wavelengths that will require experimental verification. The significance of this possible identification would be a resolution of the contradiction discussed in [20] between the current theory of carbonaceous dust formation from late stage asymptotic giant branch stars, and the observation of such an early 2175Angstrom extinction feature. If hemoglycin was responsible for the bump there would not be a contradiction in view of the very early availability of the required elements for interstellar glycine formation [11] and of Fe for hemoglycin.

In summary, hemoglycin, a homopolymer of amino acids with antiparallel $\beta$ sheet structure is known to be present in protoplanetary discs. It is in the form of 4.9nm rods that can interconnect in numerous ways to form three-dimensional lattices. Such lattices have been found to contain accreted substances such as nickel or calcium carbonate crystals. The 6.2 μm emitter is associated with bands at 7.6 and 7.8μm in a way that suggests they all might be components within very small dust grains. The ability of hemoglycin to accrete many types of material is consistent with the formation of grains. A very closely analogous antiparallel $\beta$ sheet protein, Poly-L-lysine, has its strongest infrared absorption, in the laboratory, at 1611 cm$^{-1}$ (6.207 μm). In view of these factors we propose that hemoglycin should be considered as a candidate for the source of 6.2 micron emissions. In the future, cycle 3 JWST (MIRI) spectral data for IR emission and the new 2024 Chile Vera C. Rubin Observatory instrument (LSST) for visible spectra (range 320-1060nm [21]) can be used to characterize the hemoglycin polymer.

# References

1. S. Foschino, O. Berne and C. Joblin, "Learning mid-IR emission spectra of polycyclic aromatic hydrocarbon populations from observations", *Astronomy and Astrophysics* **632**, A84 (30pp) (2019).

2. Berné, O., Martin-Drumel, MA., Schroetter, I. *et al.* Formation of the methyl cation by photochemistry in a protoplanetary disk. *Nature* **621**, 56–59 (2023). https://doi.org/10.1038/s41586-023-06307-x

3. O. Berne, private communication.

4. L. W. Beegle, T. J. Wdowiak and J. G. Harrison "Hydrogenation of polycyclic aromatic hydrocarbons as a factor affecting the cosmic 6.2 micron emission band" *Spectrochimica Acta Part A* **57**, 737-744 (2001).

5. M. Rapacioli, C. Joblin and P. Boissel "Spectroscopy of polycyclic aromatic hydrocarbons and very small grains in photodissociation regions", *Astronomy and Astrophysics* **429** 193 - 204 (2005).

6. B. T. Draine and A. Li, "Infrared Emission from Interstellar Dust. I. Stochastic Heating of Small Grains" *Astrophys. J.* **551**, 807-824 (2001). DOI 10.1086/320227

7. J. E. M. McGeoch and M. W. McGeoch. "Chiral 480nm absorption in the hemoglycin space polymer: a possible link to replication", *Scientific Reports* **12(1)** (2022) DOI: 10.1038/s41598-022-21043-4

8. J. E. M. McGeoch and M. W. McGeoch, "Polymer amide in the Allende and Murchison meteorites". *Meteoritics & Planetary Science* **50**, Nr12 1971-1983 (2015). doi: 10.1111/maps.12558

9. M. W. McGeoch, S. Dikler and J. E. M. McGeoch (2021) "Meteoritic Proteins with Glycine, Iron and Lithium" https://arxiv.org/abs/2102.10700. [physics.chem-ph] (2021).

10. N. Demirdoven, C. M. Cheatum, H. S. Chung, M. Khalil, J. Knoester and A. Tokmakoff, "Two-dimensional Infrared Spectroscopy of Antiparallel $\beta$-sheet Secondary Structure", *J. Am. Chem. Soc.* **126**, 7981-7990 (2004).

11. J. E. M. McGeoch and M. W. McGeoch, "Polymer Amide as an Early Topology". *PLoS ONE* **9**(7): e103036 (2014). doi:10.1371/journal.pone.0103036

12. C. M. O'D. Alexander, M. Fogel, H. Yabuta and G. D.Cody. "The origin and evolution of chondrites recorded in the elemental and isotopic compositions of their macromolecular organic matter". *Geochim et Cosmochim Acta* **71**, 4380-4403 (2007).

13. M. W. McGeoch, T. Šamoril, D. Zapotok and J. E. M. McGeoch "Polymer amide as a carrier of $^{15}$N in Allende and Acfer 086 meteorites". Under review *Intl. J. Astrobiology*.

14. P. Hamm, M. Lim and R. M. Hochstrasser, "Structure of the Amide I Band of Peptides Measured by Femtosecond Nonlinear-Infrared Spectroscopy", *J. Phys. Chem. B*, **102**, 6123-6138 (1998).

15. J. E. M. McGeoch and M. W. McGeoch. "Structural Organization of Space Polymers". *Physics of Fluids* **33**, 6, June 29th (2021) (DOI: 10 1063/5.0053302). https://aip.scitation.org/doi/10.1063/5.0054860

16. J. E. M. McGeoch, A. J. Frommelt, R. L. Owen, D. Lageson and M. W. McGeoch, "Fossil and present day stromatolite ooids contain a meteoritic polymer of glycine and iron", under review.

17. W.A. Schutte, K.A. van der Hucht, D.C.B. Whittet, A.C.A. Boogert, A.G.G.M. Tielens, P.W. Morris, J.M. Greenberg, P.M.Williams, E.F. van Dishoeck, J.E. Chiar, and Th. de Graauw. "ISO-SWS observations of infrared absorption bands of the diffuse interstellar medium: The 6.2μm feature of aromatic compounds." *Astron. Astrophys*. **337**, 261–274 (1998).

18. A. Li and B. T. Draine, "Infrared Emission from Interstellar Dust. II. The Diffuse Interstellar Medium", *Astrophys. J*. **554**, 778-802 (2001).

19. A. Blasberger, E. Behar, H. B. Perets, N. Brosch and A. G. G. M. Tielens, "Observational Evidence Linking Interstellar UV Absorption to PAH Molecules", *Astrophys. J.* **836**:173 (8pp) (2017).

20. J. Witstok et al., "Carbonaceous dust grains seen in the first billion years of cosmic time", *Nature* **621,** 267 – 270 (2023).

21. https://en.wikipedia.org/wiki/Vera_C._Rubin_Observatory.